# Getting the best from skylines and top-k queries

Marco Costanzo

Politecnico di Milano

Milan, Italy

marco.costanzo@mail.polimi.it

20/02/22

**Abstract**

Top-k and skylines are two important techniques that can be used to extract the best objects from a set. Both the approaches have well-known pros and cons: a quite big limitation of skyline queries is the impossibility to control the cardinality of the output and the difficulty in specifying a trade-off among attributes, whereas the ranking queries allow so. On the other hand, the usage of ranking implies that ranking functions need to be specified by users and renouncing the simplicity of skylines. Flexible/restricted skylines present a new approach to tackle this problem, combining the best characteristics of both techniques making use of a new flexible relation of dominance.

## 1 Introduction

When dealing with multi-objective optimization problems, i.e. problems in which more than one criteria need to be optimized, there are three main approaches [10] that it is worth considering: reducing the problem into a single objective using a formula involving weights (i.e. ranking functions in the case of ranking), adopting a lexicographic approach and the Pareto approach. The lexicographic approach consists in assigning different priorities to different objectives and then comparing the objects following the order of the different priorities assigned. Differently, the Pareto makes use of multi-objective algorithms. This approach permits to avoid running multiple times the same algorithm but at the price of higher complexity. Ranking and skylines are two of the main traditional approaches used to solve this type of problem, but some of their drawbacks can make their usage challenging, depending on the scenario. This survey introduces ranking (in Section 2) and skyline techniques (in Section 3) and presents a new flexible relation by analyzing also some other variations of top-k queries and skylines: OSS skylines (in Section 4), skyband queries (in Section 5), constrained skylines (in Section 6), ranked skylines (in Section 7), UTK (in Section 8), ORU/ORD (in Section 9) and $\epsilon$ - skylines (in Section 10). Then, in Section 11, flexible/restricted skylines will be introduced. For each one of the presented techniques, it will be explained how they can be used to overcome some of the limitations of top-k and skylines. The focus of this report is also on showing the applicability of these new flexible approaches and a comparison is made with respect to ranking queries and skylines. Furthermore, the applicability of all the presented techniques will be discussed with a particular focus on highlighting the advantages and drawbacks of each one of them. Finally, the conclusion (in Section 12) terminates the report and highlights the main points of the presented techniques.

# 2 Top-k queries

The result of a query to a relational database [5] is simply a set, differently from the one of a multimedia query that is a sorted list. For this reason, Garlic [6] adopted three rules to combine the results of queries in case the type returned is not the same. In particular, the three rules are conjunction, disjunction and negation rule.

**Defining top-k queries** Ranking, also called top-k, consists in finding the top k object according to a given scoring function. The scoring function is defined as a function having as domain a relation that returns a score for each tuple t of the relation and it is a function of the value of the attributes of t and of a weight vector w. For example, if points s= (1,4) , p= (2,4) and d=(3,4) are considered, the the top-2 set considering the scoring function $f(x,y) = w_1*x + w_2*y$, with $w_1$=0.5 and $w_2$=4, will contain p and d because the value of the scoring function computed of the points is: f(s)=16.5, f(p)=17 and f(d)=17.5. Since the top 2 objects are d and p, they will be the points returned. A graphical representation of the points p (in red), s (in blue) and d (in green) is the following:

Figure 1. Representation points p (in red), s (in blue) and d (in green)

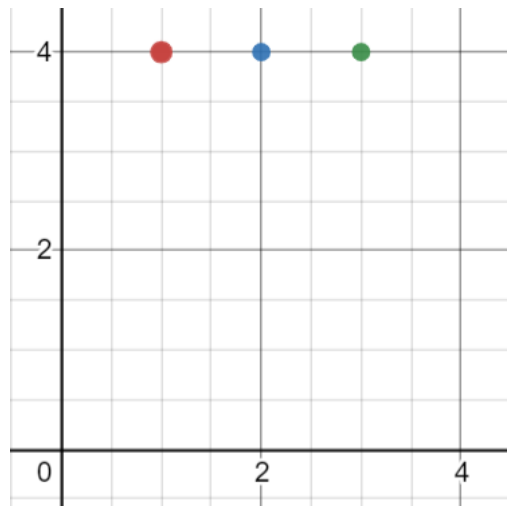

A table summarizing some ranking information is the following:

Table 1. Evaluation top-k queries

| Criteria considered | Evaluation |
|---|---|
| Allows simple formulation | No, defining the values of the weight in the scoring function in order to reflect the preferences of the user can be challenging |
| Provide an overall view of interesting results | No, because often by limiting the output size to k some relevant information will be likely missed. This problem is known as near-miss. Furthermore, it's not always easy to define the value of k. |
| Provide mechanisms for varying the cardinality | Yes, because the parameter k indicates the number of objects that will be returned in the output |
| Provide mechanisms to define a trade-off among attributes | Yes, because of the usage of different weights in the scoring function to reflect the preferences of the users |

**Multiple ranked lists** Sometimes it is necessary to obtain a single rank from several ranked lists. That ranks should be a single list [5] of objects sorted in a meaningful way.

**Distance measure between top-k lists** There are many distance measures between top-k lists [8]. In particular, it is possible to define equivalence classes of distance measures. Since a special case of a top k list is a "full list", i.e. a permutation of all the objects in the fixed universe, one possible way to define a distance measure is to compute the intersection of two top k lists. This latter distance measure gives a metric. Another approach consists in generalizing Spearman's foot rule i.e. the sum of the absolute values of the differences between the ranks in the ranking of n objects. This implies that a ranking aggregation [7] is required. In particular, it raises the need to find a ranking whose total Kendall tau distance to the given ranking is minimized. The Kendall tau distance between two rankings is the number of pairwise disagreements between the two rankings.

**Implementing top-k in SQL** To implement top-k in SQL [11] it is necessary to order the tuples (using ORDER BY, for example) and to limit the output cordiality (LIMIT or FETCH FIRST K ROWS ONLY can be used for this purpose).

> **Top-k query models** Some of the possible query models for ranking are listed below:
> 
> • **selection** In this query model scores are assumed to be associated with base tuples.
> 
> > **selection of 1 relation** When there is only a relation, two cases are possible: the input is sorted according to the scoring function or not. In the first case, reading the first k tuples is necessary, in the other case, it is possible to perform in-memory sort in case k is not too large. The latter implies that the entire input must be read so the cost is in the order of n*log(k) assuming n is the size of the dataset.
> 
> • **join** In this query model scores are assumed to be linked to the results, differently from the top-k selection query model in which scores are associated with base tuples. A possible SQL template for a top-k join query includes the join of the relations, the usage of order by according to some scoring functions and LIMIT k.
> 
> • **Top-k aggregate query model** In this query model the scores are computed for tuple groups.

## 2.1 K-NN

An approach to performing a top-k selection query consists in using the k-nearest neighbours (K-NN)[9] query so, determining the k tuples in the relation R that are closest to a point q according to a distance function d is required. The choice of d can be done among the many possible distance functions, as weighted Lp norms whose formula is the following:

Figure 2. Formula of weighted Lp norms

$$L_p(t,q) = \left(\sum_{n=1}^{m} w_i \, |t_i - q_i|^p\right)^{1/p}$$

where t and q represent the tuples of a relation r, $t_i$ and $q_i$ represent namely the value of the of the i-th attribute of the tuple t and q, m is the number of dimensions of r, $w_i$ is the weight of the i-th dimension and p is a positive number. This problem can be solved by a straightforward algorithm in a time linear in the input size.

**Applicability of top-k** The applicability of ranking depends mostly on the capability to define the value of the weights of the scoring function and the parameter k, i.e. the number of elements of the output set. In conclusion, ranking is very effective in identifying the best objects, it allows to control of the cardinality of the result very effectively but specifying the scoring function can be challenging. One of the solutions to this latter problem has been proposed [12] is using an algorithm for learning a ranking function. Computing the top-k set is computationally efficient because the complexity order is N*log(k) for unordered datasets of N elements. Furthermore, in a top-k query it is easy to express the relative importance of attributes thanks to the usage of weights. Another approach that can be used to compute the ranking consists in solving the K-NN problem. Using this approach involves shifting the problem of defining the value of the weights of the scoring function to the one of selecting an appropriate distance function. As in the original problem, it is still possible to express the relative importance of attributes but this is done by defining an appropriate distance function instead of defining the weight vector. Similarly, the formulation is not very simple because of the difficulty in selecting a suitable distance function. One way to solve the K-NN problem consists in using the algorithm L2NN [9], which makes a linear scan of all the elements of the database. This algorithm is very efficient and often doesn't explore more than the 5% of the dataset to produce significant results. The applicability of the K-NN approach depends mostly on the capability to define suitable distance function and, as the original problem of ranking, the parameter k, i.e. the number of elements of the output set.

# 3 Skylines queries

**Defining skylines** The idea of skylines is to find a good object i.e. a set of non-dominated tuples where a tuple t is said to dominate a tuple s if and only if t is nowhere worse than s and better at least once. For example, if points s= (1,4) and p= (2,4) are considered, p dominates s because the values of each of the attributes of p are greater or equal than the one of s, i.e. $2 \geq 1$ and $4 \geq 4$, and better for at least one specific dimension, i.e. for the first dimension 2>1. Consequently, p belongs to the skyline. This definition implies that a tuple t is in the skyline if and only if it is the best result w.r.t. at least one monotone scoring function. A function defined on ordered sets is defined monotone if it preserves or reverses the given order.

A graphical representation of the points s (in green) and p (in purple) is the following:

Figure 3. Representation points s (in green) and p (in purple) where p dominates s

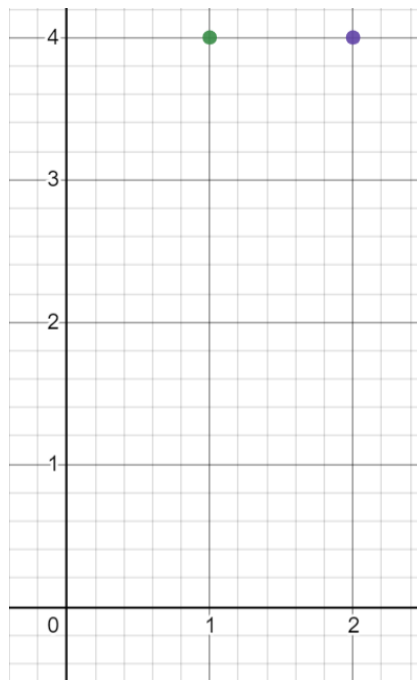

A table summarizing some information about skylines is the following:

Table 2. Evaluation of skylines

| Criteria considered | Evaluation |
|---|---|
| Allows simple formulation | Yes, because are simply defined as the set of all the tuples that are not dominated |
| Provide an overall view of interesting results | Yes, because the set of non-dominated tuples is considered interesting because they are potentially interesting results |
| Provide mechanisms for varying the cardinality | No, because skylines are not able to control the cardinality of the result set |
| Provide mechanisms to define a trade-off among attributes | No, because using skylines we can't define any trade-off among attributes |

One important property of the skyline [1] of a set M is that for any monotone scoring function from M to R, if a point p of M maximizes the scoring function, then p is in the skyline. Furthermore, for every point p in the Skyline, there exists a monotone scoring function such that p maximizes the scoring function. Generating a skyline using SQL commands is possible but not efficient: there are many algorithms with better complexity to do so.

**Applicability of skylines** The applicability of skylines is mostly limited by the impossibility to control the size of the output set[13] and to define a trade-off among the attributes of the relation considered. Skylines are easy to use and effective in domains in which there are not much information about the user's preferences.

# 4 OSS skylines

**Defining OSS skylines** One of the big limitations of skylines is the fact that they cannot control their size. OSS skylines[14], using a relation of m-domination, can overcome this limitation. A table summarizing some information about OSS skylines is the following:

Table 3. Evaluation of OSS skylines

| Criteria considered | Evaluation |
| --- | --- |
| Allows simple formulation | Yes, because are simply defined as the set of all the tuples that are not m-dominated |
| Provide an overall view of interesting results | Yes, the set of non-m-dominated tuples is considered interesting because they are potentially interesting results |
| Provide mechanisms for varying the cardinality | Yes, because varying the parameter m allows controlling the size of the OOS skyline |
| Provide mechanisms to define a trade-off among attributes | No, because using OSS skylines it is not possible to define any trade-off among attributes |

OSS skylines can be introduced by defining the notion of m-domination: a record $r_i$ m-dominates another record $r_j$ for m smaller or equal that the number of dimensions d of the relation if it does exist a set of m dimensions such that $r_i$ dominates $r_j$. This definition implies the fact that a record $r_i$ can r-dominate another record $r_j$ even if the values of its attributes are non always better than $r_j$ for all the dimensions, but are better than $r_j$ for a subset of them having a cardinality of m. Reducing the value of m corresponds to being less strict in defining the skyline that, as a consequence, reduces its size. One of the possible interpretations of m is as a measure of importance.

**Applicability of OSS skylines** The applicability of OSS skylines depends mostly on the capability to define m, i.e. the number of dimensions considered when computing the OSS skyline. Another factor that conditions their applicability is the fact that it is not possible to define a trade-off among the attributes. In conclusion, OSS skylines allow to control the size of the skyline but still do not provide a way to define a trade-off among attributes.

**Comparing OSS skylines with skylines** OSS skylines can be seen as an extension of traditional skylines, with a substantial difference: the domination relation is replaced with the m-domination relation. This choice allows OSS skylines, differently from skylines, to control their size by varying the value of m.

**Comparing OSS skylines with ranking** OSS skylines share with ranking the capability to control the cardinality of the output. In the case of ranking, this result is achieved using different weights in the scoring function to reflect the preferences of the user. control the cardinality of the output, instead, achieve this result by specifying a suitable value of m.

# 5 Skyband Query

**Defining Skyband Query** The k-skyband contains only the points that are dominated by at most K points. It can be noticed[16] that if the user scoring functions are monotonic, the top-K result is always part of the K-skyband.

**Applicability of Skyband Query** The applicability of Skyband Queries depends mostly on the definition of k, i.e. the maximum number of elements that can dominate the elements of the skyline. In conclusion, the k-skyband can be considered a 'less strict' version of the classical skyline in which the parameter k allows to specify the degree of selectivity of the result. From the practical point of view, by doing some tests [16] it was noticed that the performance of BBS degrades as K increases. This implies that in case the value of k is relatively low, this approach can be considered a valid alternative to conventional skylines.

**Comparing Skyband Query with skylines** Skyband Query can be seen as an extension of traditional skylines, with a substantial difference: the parameter k allows to specify the maximum number of points that can dominate an element of the skyline. The usage of k allows Skyband Query to be more flexible with respect to classical skylines.

# 6 Constrained Skyline

**Defining Constrained Skyline** Constrained skylines allow overcoming some limitations of classical skylines.

A table summarizing some information about Constrained Skyline is the following:

Table 4. Evaluation of Constrained Skylines

| Criteria considered | Evaluation |
| --- | --- |
| Allows simple formulation | Yes, because are simply defined as the not dominated points in the space defined by the constraints in the d-dimensional space of the attributes |
| Provide an overall view of interesting results | Yes, the set of not dominated points in the space defined by the constraints in the d-dimensional space of the attributes is considered interesting because it contains the potentially optimal results |
| Provide mechanisms for varying the cardinality | Yes, because the specification of the space in which the points of the skyline must lie using the constraints allows modifying the cardinality of the output set |
| Provide mechanisms to define a trade-off among attributes | Yes, because of the usage of the constraints to define the space in the d-dimentional space of the attributes in which the points of the skyline must lie |

Given a set of constraints, a constrained skyline[16] is defined as a skyline query whose results are the most interesting, i.e. the not dominated points in the space defined by the constraints in the d-dimensional space of the attributes.

**Applicability of Constrained Skyline** The applicability of Constrained Skyline depends mostly on the capability to define a space, using the constraints in the d-dimensional space of the attributes, that suitably model the user preferences. Constrained Skyline, differently from traditional skylines, takes into consideration only the points in the space defined by the constraints. This last property is missing in classical skylines. By doing some tests [16] it was noticed that in case the constraint region cover more than 8% of the data space, constrained skylines are more expensive, in terms of time and space than traditional skylines.

The result does not change also in case the constraint region covers almost all the data space. This result implies that it could be more advantageous to use skylines than Constrained Skyline, depending on the coverage of the constraint region.

**Comparing Constrained Skyline with skylines** Constrained Skyline Query can be seen as an extension of traditional skylines, with a substantial difference: only the points in the space defined by the constraints are considered when computing the skyline.

**Comparing Constrained Skyline with ranking** Constrained Skyline Query shares with ranking the capability to define a trade-off among attributes. In the case of ranking, this result is achieved using different weights in the scoring function to reflect the preferences of the user. Constrained Skyline, instead, achieve this result by specifying the space defined by the constraints from which the most interesting points will be taken. Similarly, Constrained Skyline, as top-k queries, can also control the cardinality of the output size
by specifying the space defined by the constraints.

# 7 Ranked Skyline

**Defining Ranked Skyline** Given a set of points in a space of d dimensions whose value of each coordinate belongs to [0,1], a parameter K, a function f defined on the attributes of the relation that is monotone on each attribute, a ranked skyline returns the K skyline points that have the minimum score according to the function f.

A table summarizing some information about Ranked Skyline is the following:

Table 5. Evaluation of Ranked Skyline

| Criteria considered | Evaluation |
|---|---|
| Allows simple formulation | Yes, because are simply defined as the set of the k tuples that have the minimum score according to the monotone function f |
| Provide an overall view of interesting results | Yes, because the set of k skyline points that have the minimum score according to the monotone function f is considered interesting |
| Provide mechanisms for varying the cardinality | Yes, because the parameter k indicates the number of objects that will be returned in the output |
| Provide mechanisms to define a trade-off among attributes | Yes, because of the usage of the monotone function f |

**Applicability of Ranked Skyline** Ranked Skyline, differently from traditional skylines, takes into input a function f and the output size k, providing more flexibility with respect to classical skylines and a simpler formulation with respect to ranking. Their applicability depends mostly on the capability to define the parameter K, i.e. the output size, and the function f that is used when computing the K skyline points.

**Comparing Ranked Skyline with skylines** Ranked Skyline can be seen as an extension of traditional skylines, with a substantial difference: the cardinality of the output set k can be controlled, and a monotone function f is taken into consideration when computing the result set. These two modification allows Ranked Skyline, differently from classical skylines, to control their size and to define a trade-off among attributes.

**Comparing Ranked Skyline with ranking** As ranking, Ranked Skyline uses a function f, i.e. a scoring function, to define a trade-off among attributes and it is possible to specify the output size. Differently from top-k, Ranked Skyline allows a simple formulation and provides interesting results.

# 8 UTK

Uncertain top-k queries [15], solve one of the problems of top-k queries: the difficulty in defining the value of the weights in the scoring function to reflect the preferences of the user. Defining the right value can be challenging because even a slight change on the wight could modify consistently the set of the top-k results.

A way to overcome this problem consists in considering the region of interest in the preference domain, i.e. an axis-parallel hyper-rectangle (a generalization of a rectangle for higher dimensions). UTK returns all possible sets containing the top-k results.

A table summarizing some information about UTK is the following:

Table 6. Evaluation of UTK

| Criteria considered | Evaluation |
| --- | --- |
| Allows simple formulation | No, defining the area of interest in order to reflect the preferences of the user can be challenging |
| Provide an overall view of interesting results | No, because, despite they return all possible top-k sets when the weight vector lies inside the region of interest, the value of k still needs to be specified. |
| Provide mechanisms for varying the cardinality | Yes, because it is possible to decide the region of interests and the value of k |
| Provide mechanisms to define a trade-off among attributes | Yes, because it is possible to define a suitable region of interest to reflect the preferences of the user |

The input of UTK are: the dataset D that is assumed to be organized using a spatial index, a positive integer k and the region of interest R that is a convex polygon inside the preference domain that has the same role of the weights in top-k queries, i.e. to reflect the user preferences. Differently from traditional top-k queries, UKT returns the top-k tuples such that the weight vector lies inside region R. Since the sum of weights over all dimensions is assumed to be 1, it is possible to shrink the domain of the weight vector to the original size minus one.

**Applicability of UTK** The applicability of UTK depends mostly on the capability to specify the output size, i.e. k, and the region of interest R, i.e. a convex polygon inside the preference domain. By doing some tests [15] it was noticed that the response time of some of the algorithms used to compute UTK increases sublinearly and remains almost constant for the largest dataset analyzed in the experiment. This property suggests that using UTK could be advantageous when the scenario presents all the conditions for its applicability. In conclusion, UTK can be viewed as a generalization of the topk approach that overcomes partially some of the drawbacks of top-k queries using the concept of area of interest to overcome the potentially problematic task of specifying the exact values of the weights in the scoring function.

**Comparing UTK with top-k queries** Since the UTK can be viewed as a generalization of the top-k approach in which multiple instances of a possible top-k set are returned when the weight vector lies inside the region of interest, the two techniques share some advantages and disadvantages. To be specific, UTK allows a simpler formulation than top-k because there is no longer the need to define the values of weights in the scoring function. On the other hand, the region of interest R needs to be specified so the problem is just shifted from computing the weights of the scoring function to defining this region. UTK provides results that are more interesting than the one of top-k queries because, since they return all possible top-k sets when the weight vector lies inside the region of interest, the result is not limited to a set of k elements only so the problem on near miss, that consists in missing some relevant information when limiting the output size to a fixed value, in partially solved. The problem of defining the exact value of k, which can be very relevant in some application, remain unsolved. As top-k, they provide a mechanism for varying the cardinality, i.e. varying the size of the region of interest and varying k. As top k, they provide a way to define a trade-off among attributes, i.e. defining a suitable region of interest that is a convex polygon inside the preference domain. As [15] reports, it is possible to arrange a top-k query in such a way as to simulate the first version of UTK. The results show that increasing the parameter k was necessary by a factor of 40 to 460 and that the output size was from 30 to 230 times the output size of the first version of UTK. This result shows that it is considering UTK instead of ranking could be advantageous depending on the scenario.

# 9 ORD/ORU

ORD and ORU technique allow overcoming some of the limitations of skylines and top-k queries. Furthermore, they enable the integration of common predicates in the framework so that is possible to specify constraints regarding the range of the value of the attributes.

A table summarizing some information regarding ORD/ORU is the following:

Table 7. Evaluation ORD/ORU

| Criteria considered | Evaluation |
| --- | --- |
| Allows simple formulation | No, because both ORU and ORD require to specify a weight vector, and determining its component can be challenging |
| Provide an overall view of interesting results | No, because both ORU and ORD require to specify the output size so some information can be missed if an inappropriate value is chosen. Furthermore, defining this value isn't always easy |
| Provide mechanisms for varying the cardinality | YES, because using both the operators it is possible to decide the value of the parameter m, i.e. the output size. |
| Provide mechanisms to define a trade-off among attributes | YES, because of the usage of different weights in the scoring function to reflect the preferences of the users |

**Defining ORU** Now ORU [14] technique will be defined: given the seed vector of weights w and the size of the output m that is required, ORU returns the records that belong to the top-k set for at least one preference vector within the distance p from w, for the minimum p that produces exactly m records in the output.

**Defining ORD** A record a p-dominates a record b if the scores of the record a are at least as high as the one of the records b for every vector v that is in the relation $|v-w| \leq p$ with the vector w, and strictly higher for at least one of them. The definition of ORD technique is the following: given the seed vector of weights w and the size of the output m that is required, ORD returns the records that are p-dominated by less than k others, for the minimum value of p that produces m records in the output.

**Applicability of ORD/ORU** The applicability of both ORU and ORU techniques depends mostly on the capability to define a weights vector to model the user preferences and the capability to identify the exact size of the output that is required. In conclusion, ORD and ORU provides a more flexible approach with respect to ranking and skyline queries and overcome some of the drawbacks of both approaches. By doing some tests [14] it was reported that ORD response time is better than ORU. Experimental results show that the differences between top-m and ORD/ORD tend to increase as m increases.

**Comparing ORU/ORD with OSS skylines** The ORD criteria of selection is very similar to the one used in the OSS skylines. To make this more clear it can be noticed that ORD returns the records that are p-dominated by less than k others for the minimum value of p that produces m records in the output and similarly the K-skyband includes only the points that are dominated by no more than K points.

**Comparing ORU/ORD with top-k queries** Since ORU returns the records that belong to the top-k set for at least one preference vector within the distance p from w, for the minimum p that produces exactly m records in the output, its relation with top-k queries is very tight.

# 10 ϵ skylines

**Defining ϵ - skylines** ϵ - skylines [17] can be defined as a generalization of conventional skylines but with a substantial difference in the definition of dominance that allows them to overcome some of the main drawbacks of classical skylines.

A table summarizing some information about ϵ - skylines is the following:

Table 8. Evaluation of ϵ - skylines

| Criteria considered | Evaluation |
| --- | --- |
| Allows simple formulation | Yes, because are simply defined as the set of all the tuples that are not ε - dominated |
| Provide an overall view of interesting results | Yes, because they inherit from skylines this feature. |
| Provide mechanisms for varying the cardinality | Yes, because varying the parameter ε allows controlling the size of the ε-skyline |
| Provide mechanisms to define a trade-off among attributes | Yes, because of the usage of different weights in the scoring function to reflect the preferences of the users |

The ϵ - dominance relation is a generalization of the conventional skyline: in particular, the two are equivalent when ϵ = 0. Given two tuples t1 and t2, t1 is said to ϵ - dominate t2 if, given a component of the weight vector wi greater than 0 and less or equal than 1 (i inside the interval [1,d]) and ϵ constant between -1 and 1 (-1 and 1 included), if for any i inside the interval [1,d], t1[i]*wi ≤t2[i]*wi+ ϵ and it does exist a j such that t1[j] <t2[j] holds with j inside the interval [1,d]. The ϵ- skyline can be greater or equal to the classical skylines, depending on the positive or negative value of ϵ. By doing some experiments[17], it was noticed that when ϵ increases, assuming ϵ greater than 0, the number of ϵ - dominated objects increases consequently. Differently, when ϵ decreases, assuming ϵ less than 0, fewer objects will be ϵ- dominated and so more objects will be retrieved because the size of the ϵ - skylines will increase. Reducing the value of ϵ when its value is negative corresponds to reducing the number of objects that will be ϵ- dominated because of the formula t1[i]*wi ≤t2[i]*wi+ ϵ in the definition of ϵ- domination. Consequently, more records will belong to the ϵ - skyline so its size will increase. Conversely, increasing the value of ϵ, assuming it is greater than 0, implies increasing the number of objects that will be ϵ- dominated because of the formula t1[i]*wi ≤t2[i]*wi+ ϵ in the definition of ϵ- domination. As a consequence, fewer records will belong to the ϵ- skyline and consequently its size will decrease.Two algorithms that can be used to compute the ϵ - skyline are ϵ - SFS and the Index-based Filter-Refinement (IFR) algorithm.

**Applicability of ϵ - skylines** The applicability of ϵ - skylines depends mostly on the capability to define the parameter ϵ that is used to control the size of the ϵ - skyline and the value of the weights that are used to compute the ϵ-dominance relation. It's interesting to notice that in case ϵ is less than 0, the algorithm ϵ - SFS run faster despite the increment of the number of objects in the ϵ - skyline. This fact should be considered when choosing the approach to adopt. ϵ - skylines have some interesting features that overcome some of the drawbacks of standard skylines, allowing to control the size of the skyline by changing the value of ϵ and integrating the weights in the formulation of the relation of dominance. Furthermore, ϵ- skylines provide a rank for all the objects.

**Comparing ϵ - skylines with top-k queries** Since both ϵ - skylines and top-k queries make use of weights to reflect the preference of the user, both approaches provide a mechanism to define a trade-off among attributes. ϵ - skylines, differently from ranking, uses the weights in the definition of the ϵ - domination relation that imposes more conditions than ranking when comparing two records. As ranking, ϵ - skylines can control the cardinality of the result set varying the value of ϵ. Differently from the parameter k of ranking, the value of ϵ is not the output size but it indirectly influences it because it is used in the definition of the ϵ - dominance relation that is used to compute the ϵ - skyline.

**Comparing ϵ - skylines with skylines** ϵ - skylines can be seen as an extension of traditional skylines, with a substantial difference: the domination relation is replaced with the ϵ-domination relation. This choice allows ϵ - skylines, differently from skyline, to control their size.

# 11 Flexible skylines

A strategy to define a trade-off among attributes in a skyline that is used by flexible skylines, also referred to as restricted skylines[3] [2] consists in imposing constraints on weights in the scoring function to reflect the preferences of the users. This specific property is missing in traditional skylines but there are many other interesting properties enjoyed by flexible skylines. One of those is the capability to provide an overall view of results that are considered interesting that is missing in ranking but that is typical of skylines. Defining different constraints implies varying the cardinality: in particular, an intuitive consequence of imposing tighter constraints is the reduction of the cardinality of the result set. Conversely, imposing less strict constraints implies the increase of the cardinality of the result set. The control of the cardinality of the result is a feature enjoyed by top-k queries, but that is missing in skyline queries. Imposing constraints on the weights to reflect the preferences of the users allows the flexible skylines to keep the formulation quite simple. The following table summarizes that information:

Table 9. Evaluation of Flexible Skylines

| Criteria considered | Evaluation |
| --- | --- |
| Allows simple formulation | Yes, because of the usage of constraints on weights to define a trade-off among attributes that is usually easier than specifying the value of the weights in the scoring function as the top-k technique does |
| Provide an overall view of interesting results | Yes, because they inherit from skylines this feature. |
| Provide mechanisms for varying the cardinality | Yes, because imposing tighter constraints or less strict constraints implicate the reduction and increase of the cardinality of the set that is the result |
| Provide mechanisms to define a trade-off among attributes | Yes, because of the usage of different weights in the scoring function to reflect the preferences of the users |

**Defining F** To deeply understand this technique the set of functions F is introduced, i.e. the set of functions defined as the weighted sum of the attributes of a relation (i.e. they are scoring functions) for which a set of constraints on their weights is defined. A simple example is shown to clarify this concept: in case space of two dimensions is considered, the set of functions belonging to F, i.e. $f_i(x,y) = w_1*x + w_2*y$ , will have to satisfy all the constraints contained in the set C. For example, if C contains the constraints: $2*w_1 > w_2$ and $3*w_1 < w_2 +3$ , then one of the functions that can belong to the set F is $f_1(x,y)=0.5*x+0.5*y$ because since $w_1$ is 0.5 and $w_2$ is 0.5, all the constraints in set C are satisfied. In this case both the first constraint $2*0.5=1 > 0.5$ and second the constraint $3*0.5=1.5 < 0.5 + 3 = 3.5$ holds.

A graphical representation of the area defined by the constraints in which the x-axis represents $w_1$ and the y-axis $w_2$ is the following:

Figure 4. The area defined by the constraints of the example

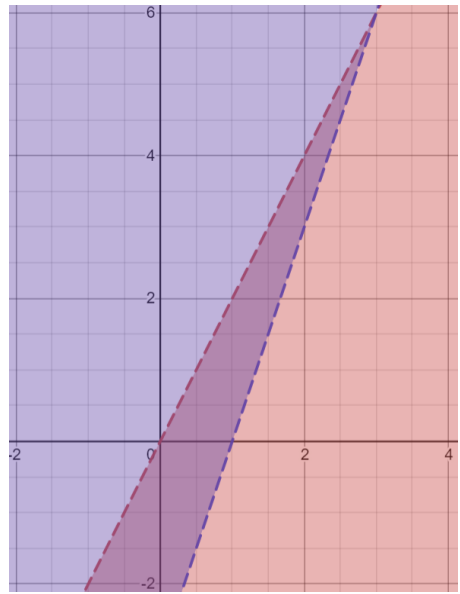

**Defining ND** To define ND, i.e. the set of non dominated tuples, the concept of F-dominance will be introduced: [4] a tuple t F-dominates another tuple s when the values obtained by computing every scoring function that belongs to F on t are always better than the one obtained by applying it on s. After defining this concept, the ND set can be defined as the set of non-F-dominated tuples. Referring to the previous example, considering the points s= (1,2) and p= (2,4), p F-dominates s because for each function contained in the set F, for example, the function f1(x,y) = 0.5*x + 0.5*y, the values of the function computed in p are greater or equal to the one of the function computed in s. In this case f1(p)=f1(2,4)=3 ≥ f1(s)=f1(1,2)=1.5. A graphical representation of the points s (in purple) and p (in green) is the following:

Figure 5. Representation points p (in purple) and s (in green) where p F-dominates s and belongs to ND

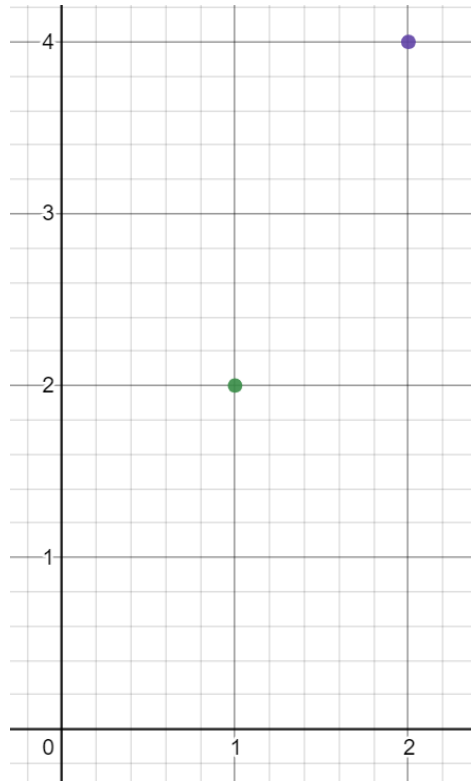

**Defining PO** Differently, PO identifies the tuples that are potentially optimal. This means that if the tuple t belongs to PO, then there exists a function in F such that f(t)>f(s) for all s different from t. This definition implies that a tuple that is PO is also in ND. Referring to the previous example, p is in the set PO because there exists a function in F, for example the function f1(x,y) = 0.5*x + 0.5*y, such that values of this function computed in p are greater than the one of the function computed in any other point. In this case, f1(p)=f1(2,4)=3 > f1(s)=f1(1,2)=1.5.

**Relations of Skylines with Flexible Skylines** The relations between the skyline and flexible skylines [3] are summarized in the following table:

Table 10. Relation between Skylines and Flexible Skylines

| Condition to impose to skylines | Relation with flexible skylines |
|---|---|
| The set of functions F coincides with the set of all the monotone scoring functions | ND and PO coincide with the skyline |
| The set of functions F contains only monotone scoring functions | PO is strictly included in ND that is strictly included in the skyline |

When computing ND and PO many algorithms can be used. To be specific, it is possible to compute ND using both linear programming and computing the F-dominance region of a tuple. The F-dominance region of a tuple s is defined as the region of the space of d dimensions, assuming d is the cardinality of the relation r for which the tuples are defined, such that if s F-dominates another tuple t, then t belong to the F-dominance region of s. Differently, the evaluation of PO is possible only using two methods both relying on linear programming. Both methods can be applied in case the set F contains only MLW functions. The set of monotonically transformed, linear-in-the-weight (MLW) functions [4] is defined by the following form:

Figure 6. Formula of monotonically transformed, linear-in-the-weight (MLW) functions

$$f^W(t) = h\left(\sum_{i=1}^{d} w_i g_i(t[A_i])\right)$$

where W is a weight vector whose components are namely wi. W belongs to the set of linear constraints and the two functions gi and h are continuous, monotone transformations such both gi and h are either nondecreasing or increasing. The set MLW cover, among other functions, the set of weighted linear functions, i.e. the same type of functions used to check the F-dominance relation. One important and quite intuitive property is that the reduction of the set of functions F implies the growth of the F-dominance region.

**Applicability of Flexible skylines** The applicability of flexible skylines depends mostly on the capability to define a suitable set of constraints on weights to model the user preferences. Defining the constraints of this set is easier than specifying the exact values of the weights of a scoring function and allows to avoid some of the problems related to this latter, such as the fact that a little variation on weights [15] can modify considerably the top-k set. Flexible skylines have some interesting features that overcome some of the drawbacks of both ranking and skylines, allowing a simple formulation but at the same time providing interesting results and controlling the cardinality of the output. Furthermore, they allow to define a trade-off among attributes.

**Comparing top-k queries with flexible skylines** After having introduced top-k queries under many aspects now a comparison is made between ranking and flexible skylines [4]. To do so two functions of the set S, that is the set containing the Skyline, ND and PO, are introduced.

**Introducing the precision function** The first function is the precision (PRE) and it is computed as the ratio between the cardinality of the set that is the intersection between S and the top-k tuples in the relation r considering f as scoring function and k. Intuitively, a high value of the precision would imply that the number of elements that are in the intersection between S and the tuples in the top-k is higher than k, so most of the elements of the top-k tuples are also present in S.

**Introducing the recall function** The second function introduced is the recall that, similarly to the precision, is defined as the ratio between the cardinality of the set that is the intersection between S and the top-k tuples in the relation r considering f as scoring function and the cardinality of the set S. A high value of the recall would imply that the number of elements that are in the intersection between S and the tuples in the top-k set is higher than the cardinality of S, consequently, most of the elements of the S are also top-k tuples. In case the skylines and top-k queries sets are not disjointed then it means that the precision and the recall of the skyline are not zero (otherwise the intersection between the skylines and top-k queries would be the empty set). In this case, since flexible skylines can be thought of as a combination of conventional skylines and top-k queries, it can be expected that the cardinality of the intersection between F-skylines and both skylines and top-k to be close as value or smaller to the cardinality of the intersection between skylines and top-k queries. Since, given a fixed input, the value of top-k depends on k and on the scoring function, the choice of this function and of k conditions the cardinality of the intersection set between top-k and skylines. Similarly, the choice of the set of constraints associated with the functions in F, the choice of k and of the scoring function associated with the top-k set conditions the cardinality of the intersection set between top-k and F-skylines. It's interesting to notice that PRE(PO)<=PRE(ND)<=PRE(SKY). This property is very similar to the property that states that in case a set of monotone scoring functions is considered: the set of PO is strictly included in the set ND, which is again strictly included in the skyline. The meaning of this chain of inequalities is that in case most of the top-k tuples are also PO, then the same will happen for ND and the skyline. Consequently, it will happen more easily that most of the top-k tuples will be elements of
the skyline rather than non F-dominated tuples and PO tuples.

**Computing PRE and REC of top-k queries** The computation of the values of the precision and the recall functions depending on k for different datasets shows the correlation between the precision computed on ND, PO and the skyline. From the definition of the functions it is easy to predict that for high values of k, the precision will tend to 0 and the recall to infinite. Comparing the complexities of the algorithms used to compute PO and ND and the one used to compute the top-k queries consistent differences can be observed for what concerns the complexity. By doing some experiments [4] it was proved that computing flexible skylines instead of top-k queries require a higher overhead. The quality of the results is, anyway, higher, i.e. flexible skylines capability to return the best objects from a set is much higher than the one of ranking thanks to their capability in allowing a simple formulation and providing interesting results. In conclusion, if a little overhead can be tolerated and defining a set of constraints on weight to model the user preferences is not prohibitively complex, flexible skylines are a valid alternative to conventional top-k queries.

**Comparing skylines queries with flexible skylines** After having introduced skylines queries under many aspects, a comparison between skylines and flexible skylines [4] is now made. To do so, the ratio between the cardinality of the set containing the points retained by the F-skyline operators, i.e. PO and ND, and the cardinality of the one containing those in the skyline will be considered.

**Computing the ratio on real datasets** The computation of the ratio on several datasets[2] shows the effectiveness of flexible skylines compared to traditional skylines. It was reported that the ratio tends to 0 when the size of the dataset grows and that PO is more effective than ND because the experiments showed that the ratio relative to ND was always higher than the one relative to PO. Another important result is the fact that when the number of constraints imposed increases, the ratio decreases, i.e. the effectiveness of ND and PO increases. This fact can be justified by observing that by imposing new constraints, the number of functions belonging to the set F that need to be considered when computing the F-skylines will reduce and so the number of tuples that are not F-dominated (i.e. the tuples belonging to the set ND) will decrease because the number of functions that will be considered when computing whether a tuple F-dominated another tuple will decrease. Considering d as the number of dimensions of the relation it was noticed that the ratio (and
in particular the one related to PO) decreases consistently as d increases. Experimental results [4] shows that the time required to compute the skyline is very close to the time required to compute ND. Differently, the computation of PO implies a moderate overhead but its effectiveness is better than the one of ND. The capability of flexible skylines in controlling the size of the output and in providing a way to define a trade-off among attributes make them a valid alternative to conventional skylines. To summarize, the experiments proved that F-skylines are always much more effective than conventional skylines and PO is let efficient than
ND but more selective.

**Comparing Flexible Skylines with Constrained Skylines** Constrained Skylines shares with flexible skylines some common ideas: the specification of the space defined by the constraint in the Constrained Skyline is very similar to the idea of specifying the user preferences using constraints on the weight of flexible skylines. To be specific, the usage of constraints on weight can be viewed as a way to specify the space from which the most interesting points will be taken when computing the Constrained Skyline because the set of functions F having a set of constraints on weight can be used to identify the F-dominance region of a specific tuple. Checking the region dominance can be used to identify PO, similarly, the space defined by the constraints is used to define the space from which the most interesting points will be taken, i.e. the result of the Constrained Skyline Query.

**Comparing Flexible Skylines with UTK** Some similarities between UTK and flexible skylines can be noticed: the specification of the region of interest as input of UTK is very similar to the idea of specifying the user preferences using constraints on weight in the latter. To be specific, the usage of constraints on weight can be viewed as a way to specify a region of interest because the set of functions F having a set of constraints on weight can be used to identify the F-dominance region of a specific tuple. Checking the region dominance can be used to identify PO; similarly, the region of interest is used to determine the output of
UTK that are all possible top-k sets when the weight vector lies inside the region of interest.

**Comparing Flexible Skylines with ORU/ORD** As flexible skylines, ORD provides a certain degree of flexibility in preference specification because it returns the records that belong to the top-k set for at least one preference vector within the distance p from w, for the minimum p that produces exactly m records in the output and not just the top-k set. Similarly, ORU also provides a certain degree of flexibility in preference specification because of the relation | v-w | ≤ p between the vector v and the vector w. This relation is used to define the relation of p-domination that is used to define the output of ORD operator as follow: the records that are p-dominated by less than k others, for the minimum value of p that produces m records in the output. The specification of the vector of weights w allows to define the preference specification in a flexible way and also a certain degree of personalization of the result varying the values of the elements of the weight vector.

## 12 Conclusions

This report presented many techniques that can be used to extract the best object from a set of objects. Each one of them has advantages and disadvantages and, depending on the type of result that is required and from the application domain, some techniques can be preferred to others. In case defining the exact number of elements of the output set and the value of weights of the scoring function is not a prohibitive task, then ranking can be considered. Differently, if no information about the user preference is available and it is not so important to define a fixed output size, then skylines could be considered. In the case in which it is possible to identify the number of dimensions that need to be considered when computing the skyline, then OSS skylines can be a valid option, considering also their capabibilty to control the output size according to the choice of the number of dimensions. In case it is possible to identify the number of points that can dominate at most an element of the skyline, then skyband queries can be considered. In case the task of specifying a set of constraints that need to be satisfied by elements of the skyline is not prohibitively complex, Constraint skylines offer a very flexible approach that is worth considering. Similarly, if it is possible to define a monotone function of the attribute of a relation and the output size, then ranked skylines should be considered, thanks also to their relatively simple formulation. If defining the region on the preference domain in which the weight vector used to rank the tuples is not difficult, then Uncertain top-k queries can be considered a valid alternative. ORU and ORD can be a valid option if, as in the case of ranking, specifying the vector of weights is not too difficult and the output size is known. In case defining the
value of $\epsilon$ of the relation of $\epsilon$ - dominance and the value of weights is not a hard task, then $\epsilon$ - skylines should be considered, also thanks to their capability to control the output size and to define a trade-off among attributes. Flexible skylines should be considered in the case in which defining constraints on the weights of the scoring function is not too challenging. Furthermore, they presented a more flexible approach with respect to ranking and skylines and offer many interesting features that overcome many of their limitations, improving by far the quality of the output while keeping a quite easy and effective formulation.

# References

[1] Stephan Borzsonyi, Donald Kossmann, and Konrad Stocker. The skyline operator. In Proceedings 17th International Conference on Data Engineering, pages 421–430, 2001.

[2] Paolo Ciaccia and Davide Martinenghi. Reconciling skyline and ranking queries. In Proceedings of the VLDB Endowment, Volume 10, pages 1454–1465, 2017.

[3] Paolo Ciaccia and Davide Martinenghi. Beyond skyline and ranking queries: Restricted skylines (extended abstract). In Proceedings of the 26th Italian Symposium on Advanced Database Systems, 2018.

[4] Paolo Ciaccia and Davide Martinenghi. Flexible skylines: Dominance for arbitrary sets of monotone functions. ACM Trans. Database Syst. 45(4), page 45, 2020.

[5] Ronald Fagin. Combining fuzzy information from multiple systems. In Proceedings of the Fifteenth ACM SIGACT-SIGMOD-SIGART Symposium on Principles of Database Systems, page 216–226, 1996.

[6] Ronald Fagin. Fuzzy queries in multimedia database systems. In Proceedings of the Seventeenth ACM SIGACT-SIGMOD-SIGART Symposium on Principles of Database Systems, page 1–10, 1998.

[7] Ronald Fagin, Ravi Kumar, Mohammad Mahdian, D. Sivakumar, and Erik Vee. Comparing and aggregating rankings with ties. In Proceedings of the Twenty-third ACM SIGACT-SIGMOD-SIGART Symposium on Principles of Database Systems, page 47–58, 2004.

[8] Ronald Fagin, Ravi Kumar, and D. Sivakumar. Comparing top k lists. in proceedings of the fourteenth annual acm-siam symposium on discrete algorithms. In Proceedings of the Fourteenth Annual ACMSIAM Symposium on Discrete Algorithms, page 28–36, 2003.

[9] Ronald Fagin, Ravi Kumar, and D. Sivakumar. Efficient similarity search and classification via rank aggregation. In Proceedings of the 2003 ACM SIGMOD International Conference on Management of Data, page 301–312, 2003.

[10] Alex Alves Freitas. A critical review of multi-objective optimization in data mining: a position paper. In SIGKDD Explorations, Volume 6, pages 77–86, 2004.

[11] Ihab F. Ilyas, George Beskales, and Mohamed A. Soliman. A survey of top-k query processing techniques in relational database systems. In ACM Computing Surveys, Volume 40, page 1–58, 2008.

[12] Thorsten Joachims. Optimizing search engines using clickthrough data. In Proceedings of the Eighth ACM SIGKDD International Conference on Knowledge Discovery and Data Mining, pages 133–142, 2002.

[13] Chengkai Li, Nan Zhang, Naeemul Hassan, Sundaresan Rajasekaran, and Gautam Das. Flexible and efficient resolution of skyline query size constraints. In IEEE Transactions on Knowledge and Data Engineering, Volume 23, pages 991–1005, 2011.

[14] Kyriakos Mouratidis, Keming Li, and Bo Tang. Marrying top-k with skyline queries: Relaxing the preference input while producing output of controllable size. In SIGMOD 2021: International Conference on Management of Data, page 1317–1330, 2021.

[15] Kyriakos Mouratidis and Bo Tang. Exact processing of uncertain top-k queries in multi-criteria settings. In Proceedings of the VLDB Endowment, Volume 11, pages 866–879, 2018.

[16] Dimitris Papadias, Yufei Tao, Greg Fu, and Bernhard Seeger. Progressive skyline computation in database systems. ACM Transactions on Database Systems, Volume 30, page 41–82, 2005.

[17] Tian Xia, Donghui Zhang, and Yufei Tao. On skylining with flexible dominance relation. In Proceedings of the 24th International Conference on Data Engineering, pages 1397–1399, 2018.